\documentclass[10pt,sigconf]{acmart}
\pdfoutput=1
\hypersetup{
  pdfauthor={Jens Finkh\"auser},
  pdftitle={Capabilities for Distributed Authorization in Information-Centric Networking},
  bookmarks=true,
  pdfpagemode=UseOutlines
}

\usepackage[english]{babel}

\renewcommand\footnotetextcopyrightpermission[1]{} 
\setcopyright{rightsretained}

\def\Interpeer/{Interpeer}
\def\InterpeerFull/{Interpeer gUG (haftungsbeschr\"ankt)}

\def\AuthServer/{\textsc{authorization server}}
\def\AuthServers/{\textsc{authorization servers}}
\def\Grant/{\textsc{authorization grant}}
\def\Grants/{\textsc{authorization grants}}
\def\Cap/{\textsc{capability}}
\def\Caps/{\textsc{capabilities}}
\def\Assert/{\textsc{assertation}}
\def\Asserts/{\textsc{assertations}}
\def\Endowment/{\textsc{endowment}}
\def\Subject/{\textsc{subject}}
\def\Person/{\textsc{person}}
\def\Persons/{\textsc{persons}}
\def\Agent/{\textsc{agent}}
\def\CredAuth/{\textsc{credential-based authentication}}
\def\CryptAuth/{\textsc{cryptographic authentication}}
\def\Id/{\textsc{identifier}}
\def\Ids/{\textsc{identifiers}}
\def\TDTC/{\textsc{time-delayed transmission channel}}
\def\AuthTuple/{\textsc{authorization tuple}}
\def\AuthTuples/{\textsc{authorization tuples}}
\def\AuthStore/{\textsc{authorization tuple store}}
\def\AuthQuery/{\textsc{authorization query}}
\def\Grantor/{\textsc{grantor}}
\def\Grantors/{\textsc{grantors}}
\def\Grantee/{\textsc{grantee}}
\def\AccessReq/{\textsc{access request}}
\def\Service/{\textsc{service}}
\def\Verifier/{\textsc{verifier}}
\def\Object/{\textsc{object}}
\def\Objects/{\textsc{objects}}
\def\Priv/{\textsc{privilege}}
\def\Privs/{\textsc{privileges}}

\acmDOI{}

\acmISBN{}

\acmConference[Interpeer Project]{}
\acmYear{2023}
\copyrightyear{2023}

\acmPrice{}

\begin{document}
\title{Capabilities for Distributed Authorization in Information-Centric Networking}
\titlenote{
  This work is licensed under the
  \href{https://creativecommons.org/licenses/by-sa/4.0/}{CreativeCommons
  Attribution-ShareAlike 4.0 International (CC BY-SA 4.0)}
  license.
}

\author{Jens Finkh\"auser}
\authornote{\url{https://orcid.org/0000-0002-1280-3658}}
\orcid{0000-0002-1280-3658}
\affiliation{%
  \institution{\InterpeerFull/}
  \streetaddress{Feldgereuth 8}
  \postcode{86926}
  \city{Greifenberg}
  \country{Germany}
}
\email{jfinkhaeuser@acm.org}

\renewcommand{\shortauthors}{Finkh\"auser}

\begin{abstract}
  Authorization currently introduces partial centralization in otherwise distributed
network architectures, such as ICN approaches. Analyzing
existing work in (partially) distributed authentication and authorization, and
rearranging proven methods, this paper introduces a generalized, capability based
and fully distributed authorization scheme.

It argues that such a scheme can fit neatly into ICN architectures in order to
enhance the trust model and mitigate against certain classes of denial-of-service
attacks.

\textit{Keywords: authorization, distributed systems security, ICN}

\end{abstract}

\ccsdesc[500]{Networks~Transport protocols}
\ccsdesc[300]{Networks~Programming interfaces}
\ccsdesc[500]{Networks~Network reliability}
\ccsdesc[500]{Networks~Network mobility}
\ccsdesc[500]{Networks~Cyber-physical networks}
\ccsdesc[500]{Networks~Security protocols}
\ccsdesc[500]{Networks~Peer-to-peer protocols}
\ccsdesc[500]{Computer systems organization~Dependable and fault-tolerant systems and networks}
\ccsdesc[300]{Computer systems organization~Fault-tolerant network topologies}
\ccsdesc[500]{Security and privacy~Authorization}
\ccsdesc[500]{Security and privacy~Distributed systems security}
\ccsdesc[300]{Security and privacy~Pseudonymity, anonymity and untraceability}
\ccsdesc[500]{Security and privacy~Access control}

\maketitle

\section{Introduction}
\label{sec:intro}

\begin{figure}[tp]
  \centering
  \includegraphics[width=\linewidth]{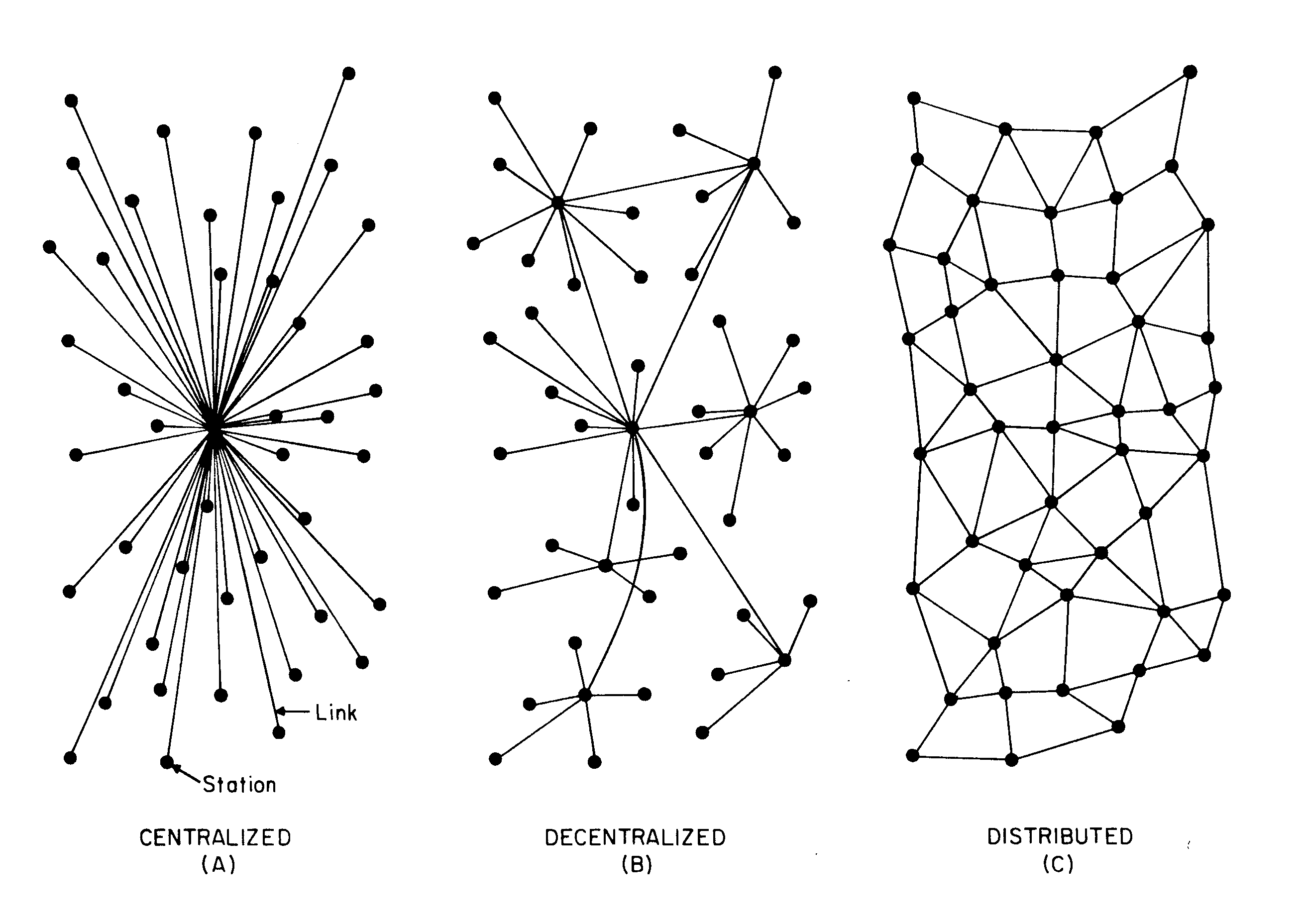}
  \caption{Centralized, Decentralized and Distributed Networks (Baran)}
  \label{fig:baran}
\end{figure}

At the advent of the Internet, Baran distinguished\cite{BARAN} between
centralized, decentralized and distributed networks. The distinguishing
characteristic lies in how many other network nodes each node maintains
connections to. In distributed networks, each node maintains multiple connections --
with the effect that a single node failure can easily be routed around (see
figure \ref{fig:baran}).

Baran was concerned with catastrophic failure; however, distributed networks
have qualities that makes them attractive for uses where node failure is part
of nominal operations, such as peer-to-peer or information-centric network
architectures.

Traditional authorization schemes in networks rely on a centralized component,
an \AuthServer/, to reply to authorization requests. This reduces distribution,
and moves the overall network architecture closer to a decentralized one.
While there is no architectural requirement to partially centralize all network
requests in this manner, authorization is a sufficiently common and critical
operation that its centralization affects a significant percentage of
requests.

This paper explores the use of cryptographic \Caps/ as a means to remove the
single point of failure that is the \AuthServer/.

The key to achieving this is to distinguish between several authorization
phases rather than treating authorization as an atomic operation. Cryptographic
signatures permit transferring state from a prior stage to a latter stage in a
verifiable and tamper-proof manner.

\section{Related Work}
\label{sec:related}

This paper combines developments from \textit{object capabilities},
\textit{cryptographic certificates} and \textit{authorization protocols}.
Most related work falls into one of those three categories. Other informative
sources revolve around \textit{distributed identifiers}.

\subsection{Object Capabilities}
\label{sec:related:ocap}

Object capabilities\cite{OCAP} are first described in a 1966 paper as a
combination of a pointer to an object, and actions that may be carried out on
that object. A key property of capabilities is that they are serializable, that
is, one process can provide them to another as part of a request for some
computation. The paper imagines a \textit{supervisor} in a multi-user machine
to arbitrate access to objects based on the capabilities a process holds.

While the paper refers to \textit{principals} as the users or
groups on whose behalf computations are performed, they are not part of the
serialized capability. Instead, the system state provides answers: if a process requests some computation
presenting a capability, it is inferred that the user who owns the process is
the principal.

Many object capability based systems are designed and explored in the meantime,
but tend to remain restricted to single domains. In 1989, Gong recognizes
that they do not translate well to distributed systems, as network user
identifiers do not necessarily map to the same local user identifier on each of
the participating nodes.

ICAP addresses this
omission\cite{ICAP}, by adding the network user identifier to the transmitted
capability. It should be noted,
however, that the \textit{access control server} is not completely removed and
remains a single point of failure in the ICAP design.

\subsection{Cryptographic Certificates}
\label{sec:related:certs}

The more popular sibling of authorization is authentication; after all, one
needs to establish an identity of sorts before deciding whether to grant them
access to some object.

Two technologies in widespread use merit mention as related work: first, X.509
certificates\cite{X509} (used in e.g. Transport Layer Security) are probably
the most popular technology that aids in identification.

The certificates contain structured metadata about an entity that, one hopes,
can be used to establish a real-world identity. An additional piece of metadata
is a cryptographic fingerprint of the public part of an asymmetric key pair.
All of this is signed using the key of some chosen authority, establishing a
\textit{security assertion}: if one can trust the authority, then one can also
trust that the identified public key belongs to the identified real-world
entity.

This shifts the issue of establishing a root in a trust chain to the distribution
of authorities, or \textit{root certificates}. In principle, applications can
choose their own root certificates. But in practice, at least as far as the world
wide web is concerned, the most commonly used root certificates are distributed as
part of operating system updates.

Where the use of X.509 certificates provides a binary answer to a trust
question, Pretty Good Privacy (PGP) provides slightly more variations\cite{rfc4880}.
In much the same way as X.509, PGP associates some identifying metadata with a
public key. Unlike X.509, however, PGP does not rely on well-known roots of
trust.

Instead, PGP opts to let the recipient of a certificate decide
whether or not to trust the signer -- and by how much. For the latter purpose,
PGP adds a piece of metadata, a numeric \textit{trust level}, which users can choose
to assign their own, locally applicable meaning to.

In this manner, PGP crosses over into the realm of authorization: by assigning
a trust level of X to a key, and then deciding to grant some privilege to all
keys with that trust level, a user effectively authorizes the affected keys.

Trust levels, however, have only locally assigned meaning. While a user may
receive a recommendation to trust a key at a certain level, both the decision
to do so, as well as the interpretation as to what that means in practice,
remains entirely under their own control. It is therefore not possible to
grant a privilege, and generate a token encoding this \Grant/ in a way that
becomes an authoritative statement.

\subsection{Authorization Protocols}
\label{sec:related:auth}

In lieu of listing the abundance of different authorization protocols in use, this paper
instead focuses only on OAuth 2.0\cite{rfc6749} as a particularly meaningful
stand-in for the class of centralized authorization protocols due to its prevalence in web
applications. Authorization flows in OAuth all require an \AuthServer/. This is in part due
to the fact that the server is also intended to authenticate clients.

OAuth generates authorization tokens that encode a \Grant/; these may be
transmitted similar to \Caps/ as part of an access request. The specific format
of token is undefined in the base specification, however. Additional
specifications\cite{rfc6750} define how such bearer tokens may be transmitted.

It is worth noting that here the specifications introduce a requirement for TLS
transport encryption, in order to mitigate against various attacks involving
the leakage of tokens. The
conclusion must then be that the specification does not expect the token to
contain any protections such as \Caps/ in ICAP above that make copying them free
of risk.

One token format that can be used is JSON Web Token (JWT)\cite{rfc7519}. JWT
are self-contained, and contain digital signatures, making them comparable in
the abstract to X.509 certificates (section \ref{sec:related:certs}).

\subsection{Distributed Identifiers}
This document focuses on authorization, which must follow authentication.
Either involves a degree of identification of the authenticated and authorized
subject. The W3C Distributed Identifier (DID)\cite{w3cDID} specification is of
some interest here, as it describes a system whereby some amount of
authentication can be achieved without requiring a server component.

This is done by a similar method as X.509 certificates, in that the DID
document may contain public key material. Unlike X.509, however, DID documents
can specify to a limited degree what purpose such public keys serve.

The
relationship between public key and additional material is also inverted: where
X.509 uses the public key as the implied identifier and associates further
material with it, DIDs are themselves identifiers. They may refer to public
keys, however, establishing a similar link.

It is necessary to stress that for this reason, DID documents are not self-verifiable.
Rather, the specification requires a verifiable storage medium.

\subsection{Seals and Notaries}
\label{sec:related:historical}

From a historical perspective, authorization has had two major forms dating
back millennia.

People recognized that rulers can not be omnipresent, so resorted to
delegation. In order to transfer to the ruler's authority to their delegates,
systems often put into places provided delegates with difficult to reproduce
seals of authority. We still rely on this mechanism in the form of e.g. police
badges or similar means of establishing sanctioned delegation of authority.

In a more generalized form, we refer to this as mechanism as "power of
attorney", a concept which can be brought into the digital realm via
cryptographic signatures \cite{PoA}.

The underlying mechanism to this is notarization, in which an established
authority signs statements that then delegate trust, establishing the rules by
which cryptographic certificates function.

\subsection{Evaluation}
\label{sec:related:eval}

While far from a complete listing of related work, the above selection serves
to highlight the common characteristics of related technologies that together
define the context in which distributed authorization must operate.

\begin{enumerate}
  \item Several technologies attempt to encode \Grants/ into
    transmittable tokens (OCAP, ICAP, OAuth, JWT).
  \item Cryptographic identification can be solved by associating some
    identifier with identifying metadata, and signing the combination with a
    trusted key (X.509, PGP, JWT).
  \item Authentication and authorization tend to be provided by an \AuthServer/
    or local service (OCAP, ICAP, OAuth).
  \item Nonetheless, some limited support for embedding \Grants/ into
    cryptographically verifiable assertions relating to identifiers (PGP, JWT,
    DID).
\end{enumerate}

As distributed authorization requires that the use of an \AuthServer/ as a
centralizing single point of failure is eliminated, the focus of this work is to
establish a generalized, flexible means of authorization using an approach
comparable to, but expanded from PGP, JWT and DID. The result is a fully self-verifiable
capability for use in distributed systems.

\section{System Design}
\label{sec:system}

When comparing the related work, it becomes apparent that it is difficult to
address authorization without first discussing authentication (section
\ref{sec:system:authentication}); to make matters
worse, authentication itself can refer to subtly different things. This paper
therefore first summarizes authentication as understood in the context of the
proposed \Cap/ scheme (section \ref{sec:system:authentication}).

Section \ref{sec:system:authorization} examines different phases of authorization in order to both
better localize how related work centralizes the authorization problem, but
also to provide the conceptual clarity required to design a distributed
authorization scheme.

Sections \ref{sec:system:validity} and \ref{sec:system:revocation} address ways
to mitigate potentially compromised keys referenced in \Caps/.

Finally, section \ref{sec:system:components} lists necessary components of a
\Cap/ in order to address all of the preceding considerations.

\subsection{(Distributed) Authentication}
\label{sec:system:authentication}

Authentication is essentially an atomic operation yielding a binary
response to the question: "Is this \Subject/ who they claim they are?"

However, the \Subject/ can be either of a real-world \Person/ or entity\footnote{
We can use "\Person/" similar to the legal sense of either a natural
person (human) or a legal person, or some other meaningful group of natural
persons.}, or it can refer to a digital \Agent/ acting on behalf of a \Person/.

This difference requires two distinct authentication processes.

\subsubsection{Authentication of a \Person/}

Authentication a \Person/ establishes their identity by comparing the
\Person/'s observable characteristics to previously established records;
thus, if e.g. their fingerprints match those on Alice's record, they are
identified as Alice.

\subsubsection{Authentication of an \Agent/}

\Agent/ identification revolves around verifying that the \Agent/ is in
possession of some secret, such as via \CredAuth/ or \CryptAuth/.

In either case, a suitable challenge/response protocol establishes whether
the \Agent/ can prove such possession.

\subsubsection{Full Trust Chain}

In order to establish that an \Agent/ acts on behalf of a \Person/, some means
of combining the two approaches above is required. In the digital realm,
observing physical \Person/ characteristic is possible by some biometric
scanning process. But in distributed systems, the \Person/ is rarely in
physical proximity of the authenticating device, requiring that authentication
becomes distributed in some way.

The method of choice, such as in TLS's use of X.509 certificates, is to
accept that there is a delay between \Person/ and \Agent/ authentication.

\Person/ authentication occurs beforehand, and endows an \Agent/ \Id/
with verifiable metadata that sufficiently proves the \Person/'s identity
for the use case. For this reason, we refer to this as \Endowment/\footnote{
In practice, automated certificate
signing via protocols such as ACME\cite{rfc8555} skip the identification of
real-world relationships, and instead focus only on whether related services
are under the requester's control. This may be deemed sufficient in some use
cases.}.

A trusted authority performs use-case relevant \Person/ authentication. It
produces a set of identifying metadata, and adds the \Agent/ \Id/. It then
notarizes the combination by adding it's cryptographic signature, producing
a cryptographic certificate.

At the point of its own authentication, the \Agent/ an then presents the
certificate along with answering the cryptographic challenge. For the trust
chain to be established, all that is required is that the \Agent/ \Id/ in
the certificate is the same as used for the challenge/response protocol (or
one can be verifiably mapped to the other).

Conceptually, this introduces a \TDTC/: between the creation of a certificate,
and the communication of it to an endpoint that needs to authenticate the
\Agent/, sufficient time passes that authentication cannot be viewed as an
atomic unit any longer.

This concept of a \TDTC/ can similarly be applied to the problem of
authorization, as described in section \ref{sec:system:authorization}.

\subsection{Distributed Authorization}
\label{sec:system:authorization}

\begin{figure*}[tp]
  \centering
  \includegraphics[width=\linewidth]{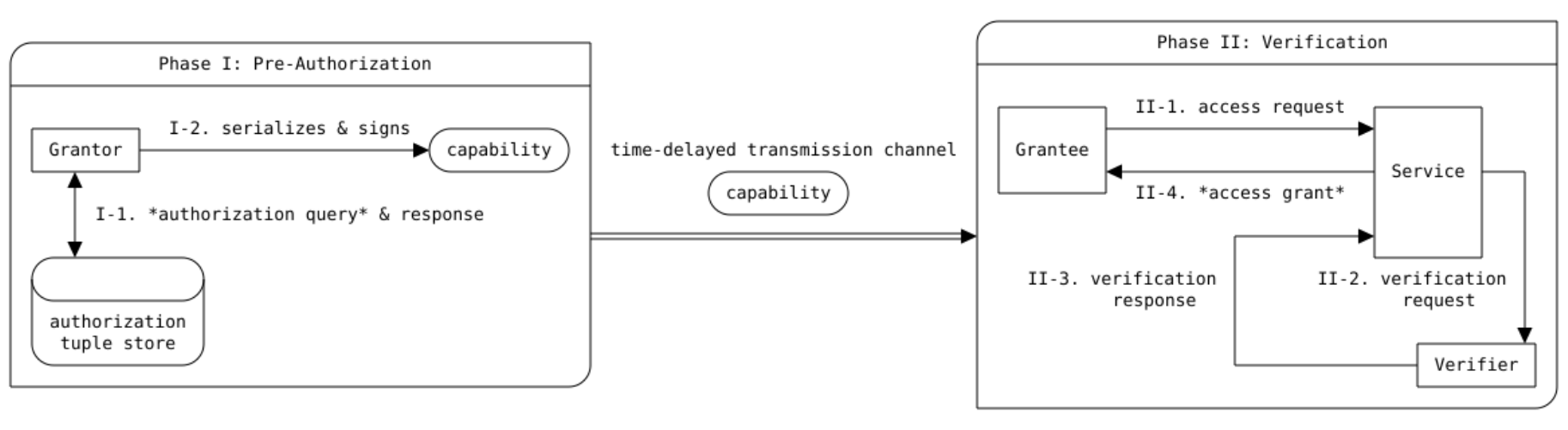}
  \caption{Architecture for capability-based distributed authorization}
  \label{fig:architecture}
\end{figure*}

In the proposed scheme, we divide the authorization process into two distinct
phases similar to the preceding analysis of distributed authentication: a
pre-authorization phase grants some privilege, and serializes it into a \Cap/.
The \Cap/ propagates to the second phase via a \TDTC/. In the second
verification phase, the \Cap/ contents are verified in a manner comparable to
that performed in authentication (section \ref{sec:system:authentication}).

Figure \ref{fig:architecture} illustrates the phases below, the components
involved and
communication flows between them.

\subsubsection{Phase I: Pre-Authorization}

The pre-authorization phase is initiated by an \AuthQuery/. A \Grantor/ with
authority to grant privileges looks up \AuthTuples/ in an \AuthStore/.

Having found an appropriate \Grant/, the \Grantor/ adds required metadata,
serializes the results and adds a signature, creating a \Cap/.

The \Cap/ enters the conceptual \TDTC/.

\subsubsection{Phase II: Verification}

The verification phase starts with an \AccessReq/ to a \Service/. The request should come from
the \Grantee/ of a \Cap/ and should include the \Cap/ in its payload.

  Strictly
speaking, the \TDTC/ can have delivered the \Cap/ independently to the
\Service/ before the request was made. However, as the purpose of capabilities
is to embed all relevant information to authorize a request, it makes sense to
treat the \Cap/ as part of the request payload. The essential part is that the
request and \Cap/ are both present at the \Service/ at this point.

It is at this point assumed that the \Agent/ is authenticated to represent the
\Grantee/, otherwise the request must fail.

The \Service/ then forwards the \Cap/ to a \Verifier/. The distinction between
\Service/ and \Verifier/ here is made purely in the functional domain; this
does not imply that they exist on separate network nodes. In fact, separating
them into distinct nodes might re-introduce a single point of failure for the
service.

The key point is that the \Service/ does not need to be aware of trust
chains, or be capable of cryptographic operations. It treats the \Verifier/ as
a function yielding a binary response: is the \Grantee/ authorized to access a
given object?

Once the \Verifier/ provides its response, the \Service/ can evaluate whether
it knows the \Object/ and supports the requested \Priv/. If that is the case,
it can grant effective access by responding to the request.

\subsection{Authorization Tuples}
\label{sec:system:tuples}

Given the above scheme, it is necessary to define the \AuthTuple/ that the
\Grantor/ looks up and signs.

The work previously done on \Caps/ is informative here (section
\ref{sec:related:ocap}): whereas OCAP defines merely a \Priv/ and an \Object/,
ICAP expands this to include a \Grantee/ as well in order to mitigate against
impersonators.

Given that the \Agent/ needs to authenticate as acting on behalf of the
\Grantee/, the ideal choice for the \Grantee/ field is either a public key that
the \Agent/ can use to authenticate, or a fingerprint or other identifier that
the \Service/ can easily map to such a public key.

Note, however, that a resolution step may introduce new system requirements that are most easily
addressed by centralized services, so care should be taken to avoid this
pitfall.

The \Object/ can be any application defined identifier. However, if the
\Object/ identifier namespace includes the \Grantee/ identifier namespace, this
opens possibilities for issuing \Caps/ on other \Persons/ in the system, which
can be used in more complex privilege management scenarios.

Similarly, the \Priv/ can be any application defined value. Existing privilege
schemes such as read/write permissions for file-like objects can be trivially
mapped into this \Cap/ scheme. However, more complex \Privs/ might include
specifying how to delegate certain \Privs/ to a different \Person/, and
establish a power of attorney scheme.

More discussion on delegation and \Persons/ as \Objects/ are outside the
scope of this paper.

Note also that these three fields can be
mapped neatly onto Resource Description Framework (RDF) N-Triples\cite{w3cRDF},
which opens the possibility of treating each field as a pointer to richer
documents. Again, however, additional resolution steps may re-centralize a
system making such a choice.

\subsection{Validity Period}
\label{sec:system:validity}

One concern with issuing storable and transmittable \Caps/ for authorization is
that a capability can be issued at a time when trust was established, but used
at a later date when trust was already lost.

Much the same issue exists in digital certificates for authentication, however,
so applying the same solution is valid: \Caps/ should include a pair of
timestamps defining the period of their validity.

Some use cases may require that validity periods are not used, however. This is
in particular the case when devices cannot have synchronized their local clock
yet, such as in e.g. Internet-of-Things (IoT) provisioning.

\subsection{Revocation}
\label{sec:system:revocation}

Issuing \Caps/ is a relatively cheap process. For this reason, validity periods
can usually be kept quite short.

However, it is always possible for trust to be lost within the validity period.
If there is then a way for the \Grantor/ to contact a \Service/ about impending
danger, that possibility should be explored.

Luckily, the data required for authoritatively granting privileges and for
authoritatively revoking them is essentially the same. It is recommended that
\Caps/ come in two flavors: grants and revocations, indicated by some flag in
their content.

In doing so, revocations similarly do not require the use of a third service
that can become a single point of failure. However, revocations must occur in
real-time, introducing such a service is unavoidable.

Adding revocations also introduces an ordering issue. Assume a grant C1 is
issued for a period of [S1, E1], then a revocation C2 is issued for [S2, E2].
Finally, a third \Cap/ C3 is issued, granting the privilege again for [S3, E3].

For simplicity's sake, assume S1 < S2 < S3, and E1 > E2 > E3; that is, the
periods are neatly nested into each other rather than overlapping.

For a given time point T, where T > S3 and T < E3, it is only possible to
determine whether the privilege is granted if the three \Caps/ are consulted in
issue order. If the consultation order were C1, C3 and C2, then C3's grant
would merely confirm C1's grant for time point, while C2's revocation would
yield the false final result.

Introducing revocations therefore introduces the need for some way to determine
the issue order of
\Caps/ -- X.509 for this purpose contains a \textsf{CertificateSerialNumber}
that issuers must strictly increment. A \Cap/ based distributed authorization
scheme should do the same.

\subsection{Capability Components}
\label{sec:system:components}

With the overall system design and additional considerations outlined in the
previous sections, it is now possible to provide a complete list of components
for \Caps/ that enable distributed authorization schemes.

\begin{description}
  \item[Grantor Identifier] \hfill \\ Due to the fact that cryptographic
    signatures are used, it is necessary to know which grantor issued a \Cap/,
    so that the signature can be verified using the appropriate key. The
    identifier here can be a full public key, or a fingerprint, or some other
    identifier that can be resolved to a public key
    (section \ref{sec:system:authorization}).
  \item[Flavour] \hfill \\ The \Cap/ can come in a grant flavor or a
    revocation flavor (section \ref{sec:system:revocation}).
  \item[Validity Period] \hfill \\ A period of validity should be provided
    using two timestamps (section \ref{sec:system:validity}).
  \item[Serial Number] \hfill \\ In order to avoid ambiguity, \Caps/ require a
    serial number (section \ref{sec:system:revocation}).
  \item[Authorization Tuple] \hfill \\ A \Cap/ must contain at least one such
    tuple. It is
    possible, however, for a \Cap/ to contain multiple \AuthTuples/ as an
    optimization (section \ref{sec:system:tuples}).
  \item[Signature] \hfill \\ Finally, a signature by the \Grantor/ over the
    preceding fields is required.
\end{description}

\section{Evaluation}
\label{sec:eval}

This paper presents a generalized concept for distributed authorization
using cryptographically signed \Caps/ as a transport mechanism over a \TDTC/.

Information-centric networks such as Named Data Networking (NDN) issue requests
in the form of Interest packets. Routers in NDN store Interests that have not
yet been satisfied in a Pending Interest Table (PIT), and forward them to other
nodes.

Interests are intended for (short-term) storage, as are \Caps/. This raises the
possibility of embedding \Caps/ in an Interest, such that data serving nodes
can reason about whether satisfying an Interest is desirable. In NDN, for
example, this could be done as part of the
\textsf{ApplicationParameters} field\cite{ndnPackets}.

The NDN trust model revolves around encryption of Data packets, such that only
parties legitimately issuing Interests should be able to make use of the
embedded payload.

While this is approach secures payload, it is not efficient. An attacker could
easily generate otherwise legitimate Interests and receive Data packets, and then discard
them. This may be used to flood caches in intermediary routers, forcing the
network to re-fetch previously stored data requested by fully legitimate
parties (cache flooding attack).

Introducing a new component to the NDN trust model is well out of scope for
this section or paper. That being said, the components and properties of the
proposed \Caps/ are sufficient for NDN routers to determine whether an Interest
should be satisfied, if

\begin{enumerate}
  \item the \Grantee/ can be mapped to the \textsf{KeyLocator} element of an
    Interest,
  \item the \Priv/ is chosen appropriately, and
  \item the \Object/ can be mapped to the \textsf{Name} element of the
    Interest.
\end{enumerate}

\section{Conclusion}
\label{sec:conclusion}

\Caps/ for distributed authorization fit well into existing information-centric
network designs, and can serve as a component in enhancing the network's trust
model in order to mitigate against cache flooding attacks.

By decoupling cryptographic authorization into two distinct phases, \Caps/
become storable without losing authority,
and thus easily embeddable in Interest packets.

Further research is required, however, to present a full enhancement to the
trust model. In particular, routers need to become aware of whether \Grantors/
are authoritative for a requested \textsf{Name} without requesting and storing the
associated Data packet.



\begin{acks}
This work has received funding from the \grantsponsor{isoc}{Internet Society Foundation}{https://isocfoundation.org} as part of their Research Program.
\end{acks}

\bibliographystyle{ACM-Reference-Format}
\bibliography{reference}

\end{document}